\documentclass{article}
\usepackage{microtype}
\usepackage{graphicx}
\usepackage{subfigure}
\usepackage{booktabs} 
\usepackage{amsmath,amssymb}

\usepackage{hyperref}

\usepackage[accepted]{icml2019}

\icmltitlerunning{Submission and Formatting Instructions for ICML 2019}

\begin{document}
\twocolumn[
\icmltitle{ CrossoverScheduler: Overlapping Multiple Distributed Training Applications in a Crossover Manner }




\begin{icmlauthorlist}
\icmlauthor{Cheng Luo}{to}
\icmlauthor{Lei Qu}{goo}
\icmlauthor{Youshan Miao}{goo}
\icmlauthor{Peng Cheng}{goo}
\icmlauthor{Yongqiang Xiong}{goo}\\

\icmlaff{State Key Laboratory of ASIC and System, Fudan University}{to}\\
\icmlaff{Microsoft Research}{goo}\\

\end{icmlauthorlist}

\icmlaffiliation{to}{Department of Computation, University of Torontoland, Torontoland, Canada}
\icmlaffiliation{goo}{Googol ShallowMind, New London, Michigan, USA}

\icmlcorrespondingauthor{Cieua Vvvvv}{c.vvvvv@googol.com}
\icmlcorrespondingauthor{Eee Pppp}{ep@eden.co.uk}

\icmlkeywords{Machine Learning, ICML}

\vskip 0.3in
]



\printAffiliationsAndNotice{}  

\begin{abstract}
Distributed deep learning workloads include throughput-intensive training tasks on the GPU clusters, where the Distributed Stochastic Gradient Descent (SGD) incurs significant communication delays after backward propagation, forces workers to wait for the gradient synchronization via a centralized parameter server or directly in decentralized workers. We present CrossoverScheduler, an algorithm that enables communication cycles of a distributed training application to be filled by other applications through pipelining communication and computation. With CrossoverScheduler, the running performance of distributed training can be significantly improved without sacrificing convergence rate and network accuracy. We achieve so by introducing Crossover Synchronization which allows multiple distributed deep learning applications to time-share the same GPU alternately. The prototype of CrossoverScheduler is built and integrated with Horovod. Experiments on a variety of distributed tasks show that CrossoverScheduler achieves $20\% \times$ speedup for image classification tasks on ImageNet dataset. 
\end{abstract}


\section{Introduction}
\label{submission}

With machine learning models becoming popular in various demanding domains \cite{Russakovsky2015ImageNet,ren2015faster,jia2014learning}, methods that can efficiently train machine learning models have attracted more and more attention recently. Due to the limited computing resource of a single machine, distributed training has been viewed as the primary method. The current widely used approach of distributed training is data parallelism, in which each worker keeps a replica of the whole model, processes training samples independently, and synchronizes the gradients every iteration. Gradient synchronization is involved in this process, usually using the parameter server architecture \cite{dean2012large} or all-reduce based gradient aggregations~\cite{hannun2014deepspeech}.

Unfortunately, the performance of distributed training is far from linear speed-up, due mainly to the communication overhead for gradient synchronization among all workers. Since the computing power of computational units grows much faster than the growth of network bandwidth, network communication performance has now become the training bottleneck, especially when the communication/computation ratio is high \cite{alistarh2017qsgd}.

Consequently, many different communication acceleration approaches have been proposed to mitigate communication overhead. One approach can increase computation-to-communication ratio by allowing workers to perform local gradients accumulation instead of global gradient synchronization \cite{su2015experiments,zhang2016parallel} or reduce the required data communications in each iteration by sending compressed gradients \cite{alistarh2017qsgd,lim20183lc,tang2019doublesqueeze}. On the other hand, Overlap communication with computation by scheduling the order communication operations can effecitvely minimize the training time \cite{hashemi2018tictac, bao2020preemptive}.

Ideally, there should be a third way to improve resource utilization and minimize the training time, that is packing multiple distributed deep learning applications to the same servers \cite{bai2020pipeswitch}. Therefore, These multiple independent distributed allocations can interfere with each other through their shared GPUs and network. However, even without over-subscription of resources, co-located distributed jobs on the same server may interfere with each other negatively and experience performance unpredictability. This is because the application share underlying resources such as GPU utilization and network I/O.

To tackle these problems, in this paper we study a communication scheduling method called CrossoverScheduler, which aims to overlap computation and gradient transmission between different distributed training applications via time-sharing.
The contribution of this paper can be summarized as follows:


\begin{itemize}
\item  CrossoverScheduler enables running computation and communication of different distributed training applications models in an crossover manner. We could ensure that the delay of gradient synchronization can be removed when the communication/computation ratio is less than $1.0$.
\item Gradients fusion is introduced to bucket all gradient into one group for synchronization, which aims to minimize communication overhead and maximize communication/computation overlap.
\item We conduct experiments on both parameter server architecture and allreduce architecture under different scales. The results show that our proposed CrossoverScheduler can achieve $10\% ~ 20\%$ speedup for all situations.
\end{itemize}


\section{Design}
In this section, we have a detailed discussion of the CrossoverScheduler. We firstly elaborate the design principle of CrossoverScheduler and then specify the algorithm flow.

\subsection{Crossover Synchronization}
The updating process of distributed training has two steps: computing gradients in each worker and gradients synchronization through network. Traditional distributed training task needs to wait for the completion of gradients synchronization before beginning the computation of next iteration. Although the gradients synchronization can be partially overlapped with backward propagation \cite{zhang2017poseidon} or the forward propagation of the next iteration \cite{peng2019generic}, it is not enough to cover communication cost since the synchronization step suffers from a global barrier or system overhead.

\begin{figure}[t]
	\centering
	\includegraphics[width=0.9\linewidth]{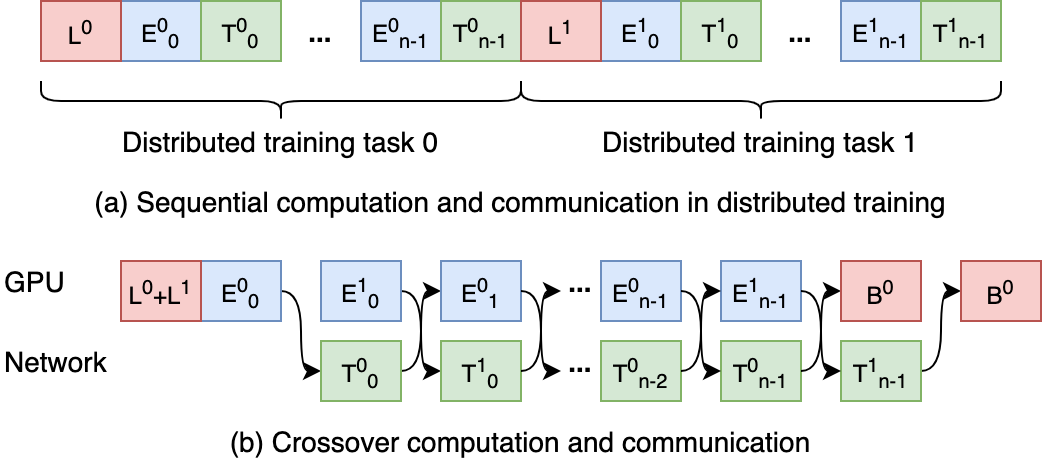}
	\caption{CrossoverScheduler pipeline task execution and model transimission}
	\label{fig:CrossoverScheduler}
\end{figure}

Figure \ref{fig:CrossoverScheduler} illustrates the advantage of CrossoverScheduler over the traditional solution where two steps could de-facto be parallel as well such that the huge delay incurred by synchronization could be totally diminished. Each worker creates a thread which is responsible for two distributed training tasks. In the beginning,  gradients are computed based on the local model and send for synchronization. It proceeds to compute the other model without waiting for completion of gradients synchronization. Then, the gradients synchronization of the first task and the gradients computation are processed in parallel until both of them are completed. CrossoverScheduler alternates the communication and computation process of the two tasks to the end.

Noted that tensor fusion is introduced after gradient computation is finished, where all tensors from one model are fused together to form a united tensor before sending this fused tensor across the network. 
In this manner, all gradients are transferred simultaneously to reduce latency overhead, and the gradients synchronization can be completely overlapped with both the forward and backward propagation of next distributed task.

\subsection{Algorithm Description}

The pseudocode of CrossoverScheduler is shown in Algorithm. CrossoverScheduler makes each GPU perform gradients computation and gradients synchronization in parallel with different models. This algorithm can be applied on  both parameter-server architecture and allreduce architecture.

Noted for the first iteration $t=1$, (lines 7--8) of Algorithm 1 is bypassed as there is gradients waiting for synchronization. Similarly, after all iterations are completed, there will be an additional gradient synchronization phase.

\begin{algorithm}[tb]
   \caption{CrossoverScheduler}
   \label{alg:example}
\begin{algorithmic}[1]
   \STATE {\bfseries Input:} Initialize model parameters $x_i$, and number of total iterations $T$
   \FOR{iteration $t \in {1,2,...,T}$}
    \FOR{sequential distributed task i $\in {1,2,...,N}$}
       \STATE Gradient computation $\nabla g^t_i \gets \nabla F(x^t_i,\varsigma)$
       \STATE Fuse all gradient into one group
       \STATE Send $\nabla g^{t}_i$ for gradient synchronization
       \STATE Receive averaged gradients  $\nabla g^{t-1}_{i+1}$
       \STATE Model Update $x^t_{i+1} \gets SGD(x^{t-1}_{i+1}, g^{t-1}_{i+1})$ 
    \ENDFOR
   \ENDFOR
   
\end{algorithmic}
\end{algorithm}


\section{Evaluations}

In this section, we conduct experiments that compared CrossoverScheduler with other distributed training strategies. Experiments are running on both the parameter server architecture and allreduce architecture, and show that, CrossoverScheduler converges similar to SGD but runs much faster.

\subsection{Experiment Setup}
We evaluate our CrossoverScheduler on image classification task by training ResNet-50 an VGG16 on imagenet. The batch size is set as 32 on each wokrer. We make evaluations on two servers, each with 2 Xeon CPU E5-2690 and 8 Tesla P100 GPUs, connected by 10Gbps Ethernet and 100Gbps InfiniBand. We implement all distributed training tasks on PyTorch \cite{li2020pytorch} and adapt Horovod~\cite{sergeev2018horovod} and BytePS~\cite{peng2019generic} as the distributed training backend.

\subsection{Results of CrossoverScheduler}

\begin{figure}[t]
\subfigure{
\begin{minipage}[t]{0.45\textwidth}
\centering
\includegraphics[width=\textwidth]{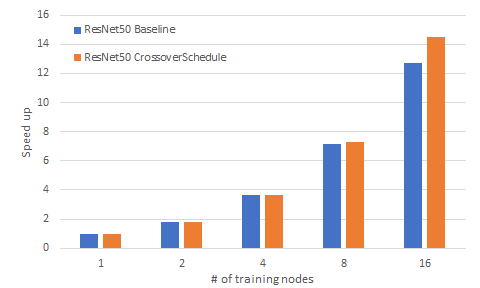}
\end{minipage}
\vspace{-0.5in}
}
\subfigure{
\begin{minipage}[t]{0.45\textwidth}
\centering
\includegraphics[width=\textwidth]{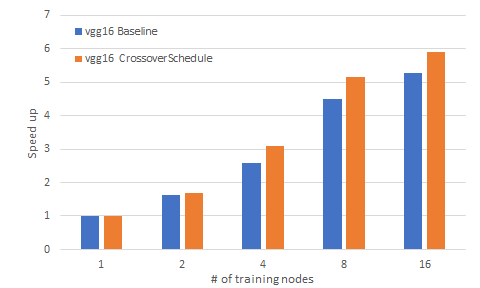}
\end{minipage}
\vspace{-0.5in}
}

\caption{CrossoverScheduler improves the speedup and scalability of distributed training.}
\label{fig:experiment}
\vspace{-0.2in}

\end{figure}

Figure \ref{fig:experiment} shows the performance and scalability for CrossoverScheduler compared with single-node training under 100Gbps InfiniBand with the number of GPUs ranging from 1 to 16. Here each machine has 8 GPUs, the experiment on 16 GPUs is employed on two servers. We see that CrossoverScheduler outperforms baseline by $10\% - 20\%$ across the 3 benchmark models. 

\section{Conclusion}
In this paper, we study a scheduling algorithm, namely CrossoverScheduler that overlaps computation and communication time of multiple distributed training task. As a result, this algorithm can significantly improve the GPU utilization and training throughput while maintain same convergence rate compared with original distributed training algorithm. 

\nocite{langley00}

\bibliography{example_paper}
\bibliographystyle{icml2019}





\end{document}